\documentclass[12pt]{iopart}
\usepackage {enumerate}
\usepackage{tikz}
\usepackage{multirow}
\usepackage{float}
\usepackage{graphicx}
\usepackage{subfigure}
\usepackage{cite}

\begin{document}

\title{Study of Parametric Instability of gravitational wave detectors using silicon test masses}
\author{Jue Zhang, Chunnong Zhao, Li Ju, David Blair}
\address{School of Physics, The University of Western Australia, Crawley, WA 6009, Australia}
\vspace{10pt}
\begin{indented}
\item[]09/09/2016
\end{indented}

\begin{abstract}
 
Parametric instability is an intrinsic risk in high power laser interferometer gravitational wave detectors,  in which the optical cavity modes interact  with the acoustic modes of the mirrors leading to exponential growth of the acoustic vibration. In this paper, we  investigate the potential parametric instability for a proposed next  generation gravitational wave detector based on cooled silicon test masses. It is shown that  there would be  about 2 unstable modes per test mass, with the highest parametric gain of $\sim 76$.  The importance of developing suitable instability suppression schemes is emphasized.
\end{abstract}
\section{Introduction}
 
Gravitational waves, predicted by Einstein a century ago, have finally been directly detected \cite{abbott2016observation} in 2015, enabling us to have a first glimpse of two black holes merging.  The detections \cite{abbott2016observation}  \cite{abbott2016gw151226} gravitational waves open the new windows on the universe and there are now strong reasons for improving detector sensitivity.  At their design sensitivity,  the kilometer scale interferometer detectors such as Advanced LIGO detectors \cite{aasi2015advanced} and Virgo detector \cite{accadia2012virgo} are expected to be able to detect $\sim10^3$ gravitational waves events per year \cite{belczynski2016first}.
 This data will be a critical resource  for understanding the origin and distribution of stellar mass black holes in the universe. 
 
To reach high sensitivity, advanced laser interferometer detectors must use very high laser power inside the optical cavities to reduce quantum noise.  Braginsky pointed out   \cite{braginsky2001parametric} in the early 2000s that with extremely high laser power in the  interferometer cavity, radiation pressure coupling between the acoustic modes and cavity optical modes can cause 3-mode parametric instability (PI), resulting in the exponential growth of many acoustic mode amplitudes and disruption of the operation of the interferometer.  Zhao et al.\cite{zhao2005parametric} undertook a detailed  3D study of  PI in a detector similar to the advanced LIGO design. The modeling considered interactions in individual arm cavities, and  made predictions \cite{zhao2005parametric} that were in relatively close agreement to recent observations of PI on the Advanced LIGO detectors \cite{PhysRevLett.114.161102}. Zhao$'$s paper discussed the use of small variations of the mirror radius of curvature to tune the system to a local minimum of parametric gain. In fused silica the thermal distortion of test masses is used for this purpose. Subsequent detailed modeling \cite{gras2010parametric} 
showed that interactions between arm cavities, power recycling and signal recycling cavities could significantly modify the predictions, and further work identified many methods of suppressing PI, such as electrostatic feedback, optical feedback and the use of passive dampers \cite{degallaix2007thermal,ju2008strategies,gras2008test,gras2009suppression,fan2010testing,miller2011damping,zhao2015parametric,Gras2015damper}. 
Unfortunately the conditions that enable high sensitivity appear to always coincide with the conditions that enable PI. To date, no method has been demonstrated that can completely eliminate parametric instability.  As advanced LIGO power is increased, it is anticipated that a combination of of thermal tuning, electrostatic feedback and passive dampers will be able to eliminate all instabilities.

To increase the sensitivity of detectors beyond that achievable with the current advance LIGO detectors, various options are being investigated.  Much effort is underway to design the next generation of detectors.  Proposed designs include the Einstein Telescope (ET) \cite{punturo2010einstein} , several LIGO upgrades designs such as LIGO A+, LIGO Voyager \cite{instruments}, a 40km interferometer \cite{dwyer2015gravitational}
and an 8km interferometer\cite{blair2015next}.  
All the future designs required the use of high laser power to overcome quantum shot noise for high frequency sensitivity.  It is apparent that each of these designs will need to consider the PI problem. In 2012, Strigin studied the PI problem of the 10km Fabry-Parrot cavity in ET design with sapphire or silicon test masses \cite{strigin2012effect}.  His result  suggested that there would be $\sim 10$ unstable modes in such systems. 

Parametric instability arises through the coincidental matching of test  mass acoustic mode frequencies and mode shapes  with optical cavity transverse modes for which the frequency difference from the main TEM$_{00}$ pump laser mode is in the 1 - 100kHz range.  Detailed predictions must take into account the details of the test mass shapes and optical cavity design.  However in general, the risk of modal coincidences depends on the optical modes density and the acoustic mode density. The optical mode density depends on the cavity free spectral range, determined by the cavity length, and the transverse mode spacing called the mode gap, which is proportional to the inverse cosine of the cavity g-factor. The acoustic mode density depends on test mass sizes and inversely on the sound velocity.  
The particular design analyzed here replaces the 40kg fused silica test masses of Advanced LIGO with $\sim$ 200kg cooled silicon test masses.  The high sound velocity and low acoustic loss of silicon could enable significant suppression of thermal noise which can be further improved through use of crystalline optical coatings\cite{cole2013tenfold}.  
The high sound velocity of silicon acts to reduce modal coincidences that lead to PI, but the increased mass acts in the other direction. 

This paper investigates this issue in detail for the specific design called the LIGO voyager blue design.  While the results are specific to this design, they give an indication of the problems that could be faced by many future  $3^{rd}$ generation gravitational wave detectors.  
For this analysis we use a single cavity model for PI modeling because of its general agreement with observations. Any more complex models depend on details of the power and signal recycling cavities, and do not change the general pattern of instability.  Hence this model provides a good indication of the level of the PI.     

 

\section{Parametric instability}
	\subsection{Parametric instability}
 
    The 3-mode parametric interaction involves 2 optical modes (the carrier fundamental mode $TEM_{00}$ and the higher order optical mode $TEM_{mn}$) and an acoustic mode.  If the transverse modes and acoustic modes have a similar spatial distribution and appropriate frequency relations, three mode interactions can be strong. The frequency match between carrier mode $\omega_{00}$, transverse mode $\omega_{mn}$ and acoustic mode $\omega_{m}$ is represented by the three-mode detuning parameter
    \begin{equation}
    \triangle \omega=|\omega_{00}-\omega_{mn}|-\omega_{m}.
    \label{eq df}
    \end{equation}
When $\triangle \omega < \delta$, where $\delta$ is the line-width of transverse mode, strong interaction will occur. 
    
    The parametric gain $R$ is used to describe the three-mode interaction in the cavity. $R$ can be described as follows in the case of single optical cavity: 
    \begin{equation}
    R=\frac{4P Q_{m}Q}{McL\omega_{m}^2}\left(\sum_{j}^{TEM}\frac{\Lambda_{s}}{1+\frac{\triangle\omega_{s}^{2}}{\delta_{s}^{2}}}-\sum_{j}^{TEM}\frac{\Lambda_{a}}{1+\frac{\triangle\omega_{a}^{2}}{\delta_{a}^{2}}}\right),
\label{R equation}
\end{equation}
where $P$ is the circulating power in the cavity arm, $\Lambda$ is the overlapping factor, $Q_{m}$ is the quality factor of the acoustic mode, $Q$ is the quality factor of the optical mode, $M$ is the mass of the test mass, $c$ is the velocity of light, $L$ is the length of the cavity, $\omega_{m}$ is the frequency of acoustic mode, $\triangle \omega_{s/a} \equiv \omega_{0/mn}-\omega_{mn/0}-\omega_{m} $ is the frequency detuning parameter of Eq.~\ref{eq df}, describing either excitation or damping of the acoustic modes, $\delta$ is the half line-width of the optical mode.

In the Stokes process ($\omega_{mn} < \omega_{00}$), the optical energy is transferred to acoustic mode, exciting the acoustic mode. For $R>1$, the exciting force exceeds damping force, the acoustic mode amplitude will rise exponentially leading to instability; $0 < R < 1$, the amplitude will be amplified by a factor $\sim (1-R)^{-1}$ and the system will remain stable. In the Anti-Stokes process ($\omega_{mn} > \omega_{00}$), $R<0$, the acoustic mode transfer energy to the optical mode, and the acoustic mode is cold damped.  Stocks and anti-stocks processes could all happen in an optical cavity, but due to the asymmetrical structure of the Stokes and anti-Stokes process, they cannot be canceled out\cite{braginsky2001parametric}.
Moreover, multiple optical modes can interact with one acoustic mode simultaneously \cite{ju2006multiple}, which is reflected by the summation of Eq.~\ref{R equation}.    
   

From eq. \ref{R equation}, it can be seen that $R$ depends on frequency match between the acoustic mode frequency and the cavity mode frequency $\Delta\omega_{s/a}$, as well as the overlap $\Lambda_{s/a}$ between the acoustic modes and the optical cavity mode.  These will be addressed below in detail for a particular interferometer design.  
  
    \subsection{LIGO Voyager design}
    
	 The  Voyager conceptual design is proposed as a major upgrade of Advanced LIGO. It was first proposed in 2012 and improved based on simulations and experiments. Subsequent studies of  Voyager led to 3 "straw-man" designs, known as  Red, Green and Blue. Here the Blue design is used for our simulation. The blue design makes feature of \cite{VoyagerDesign}:
    \begin{enumerate} [(1)]
    
    \item laser wavelength of 2000 nm
    \item Large cryogenic Silicon mirror:
204 kg Silicon mirrors at the operating temperature of 123 K are considered.
\item Silicon cryogenic suspension:
Silicon ribbons (and perhaps blade springs) in the
final stage suspension at 123 K are employed.
\item Newtonian noise:
Newtonian noise is subtracted by a factor of 30 with seismometer arrays.
\item High power 2000 nm laser:
2000 nm wavelength lasers operating at $300\sim 700$ W are
employed. Arm cavity powers will reach $2\sim 5$ MW
\item Coating Thermal Noise:
The beam spot size is kept at the aLIGO size level to avoid
optical degeneracy issues and epitaxial coatings on the test masses are employed for a mirror thermal noise improvement of 3 to 10.
\item Quantum noise:
Squeezed light injection (10 dB) and a 300 m filter cavity for frequency dependent squeeze angle is assumed.
    \end{enumerate}
    
 The parameters used for calculating parametric gain for the Voyager Blue design are listed in Table \ref{properties}.        
 
  	\begin{table}[H]    
    \centering
  	\begin{tabular}{c c c }
  	  
  	  \hline
    Parameter & Advanced LIGO & Voyager Blue \\
    \hline 
    Mirror substrate & fused silica& silicon\\
    Young's modulus&73 GPa&189 GPa\\
    Poisson ratio&0.167&0.181\\
    Density&2220 $kg/cm^{3}$&2332 $kg/cm^{3}$\\
    
    Mirror radius & 0.17m & 0.225m \\
    
    Mirror thickness& 0.20m &0.55m\\
    
    Beam radius on ITM/ETM &0.053m/0.062m &0.059m/0.084m\\
    
    Mirror mass&40kg&204kg\\
    
    Final Stage temperature&300k&123k\\
    
    Transimissitivity of ITM &$1.4\%$&$1.4\%$\\
    
    Acoustic mode quality factor & $1\times 10^{7}$ & $1\times 10^{7}$\\
    
    radius of curvature of ITM/ETM & 1934m/2245m & 1801m/2596m\\
     
    Cavity length&3994.5m&4000m\\
    
    Finesse&450&450\\
    
    wavelength&1064nm&2000nm\\
    
    Circulating Power&830 kW&3 MW\\

    \hline
    \end{tabular}
    \caption{Main parameters used in this paper compared to Advanced LIGO}
    \label{properties}
    \end{table}
    
  	\subsection{Test mass acoustic modes} 
  
                We used the finite element analysis software COMSOL to simulate the acoustic modes of the test mass using the properties shown in table \ref{properties}, from $\sim$ 5kHz up to $\sim 74$ kHz, in total 2194 acoustic modes. Compared with aLIGO test mass, the mode density is of comparable magnitude.

For the Voyager blue design, it is proposed \cite{VoyagerDesign} to have black coating on the test mass barrel for radiative cooling. It is clear that the lossy black coating on the barrel will reduce the Q-factor of acoustic modes and increase the thermal noise \cite{gras2009suppression}, depending on their strain energy distribution on the barrel. We assume the acoustic mode Q-factors are reduced by an order of magnitude from the silicon intrinsic Q factor of $10^8$ to $10^7$ due to the combined effects of both black coating on the barrel and the optical coating on the surfaces.

                
    \subsection{Optical cavity modes}
       The frequencies and shapes of high order modes also need to be determined to calculate the overlap with certain acoustic modes. The higher order Hermite-Gaussian mode with transverse mode number (n,m) and longitude number k can be written as \begin{equation}
                u_{nmp}(x,y,z)=u_{n}(x,z)u_{m}(y,z)e^{(ikz-\frac{ik\pi}{2})},
                \end{equation} 
                where 
                \begin{equation}        
                u_{n}(x,z) =\left(\frac{2}{\pi}\right)^{\frac{1}{4}}\left(\frac{e^{i(2n+1)\phi(z)}}{2^{n}n!\omega(z)}\right)^{\frac{1}{2}}\times  H_{n}\left(\frac{\sqrt{2}x}{\omega(z)}\right)exp\left[-i\frac{kx^{2}}{2R(z)}-\frac{x^{2}}{\omega^{2}(x)}\right]
                \end{equation}
$H_{n}(x)$ are the Hermite polynomials of order n, and
                \begin{equation}
                H_{0}(x)=1, H_{n+1}=2xH_{n}(x)-2nH_{n-1}(x).
                \end{equation}
                
\begin{figure}[H]
				\centering
				\includegraphics[width=4.7in]{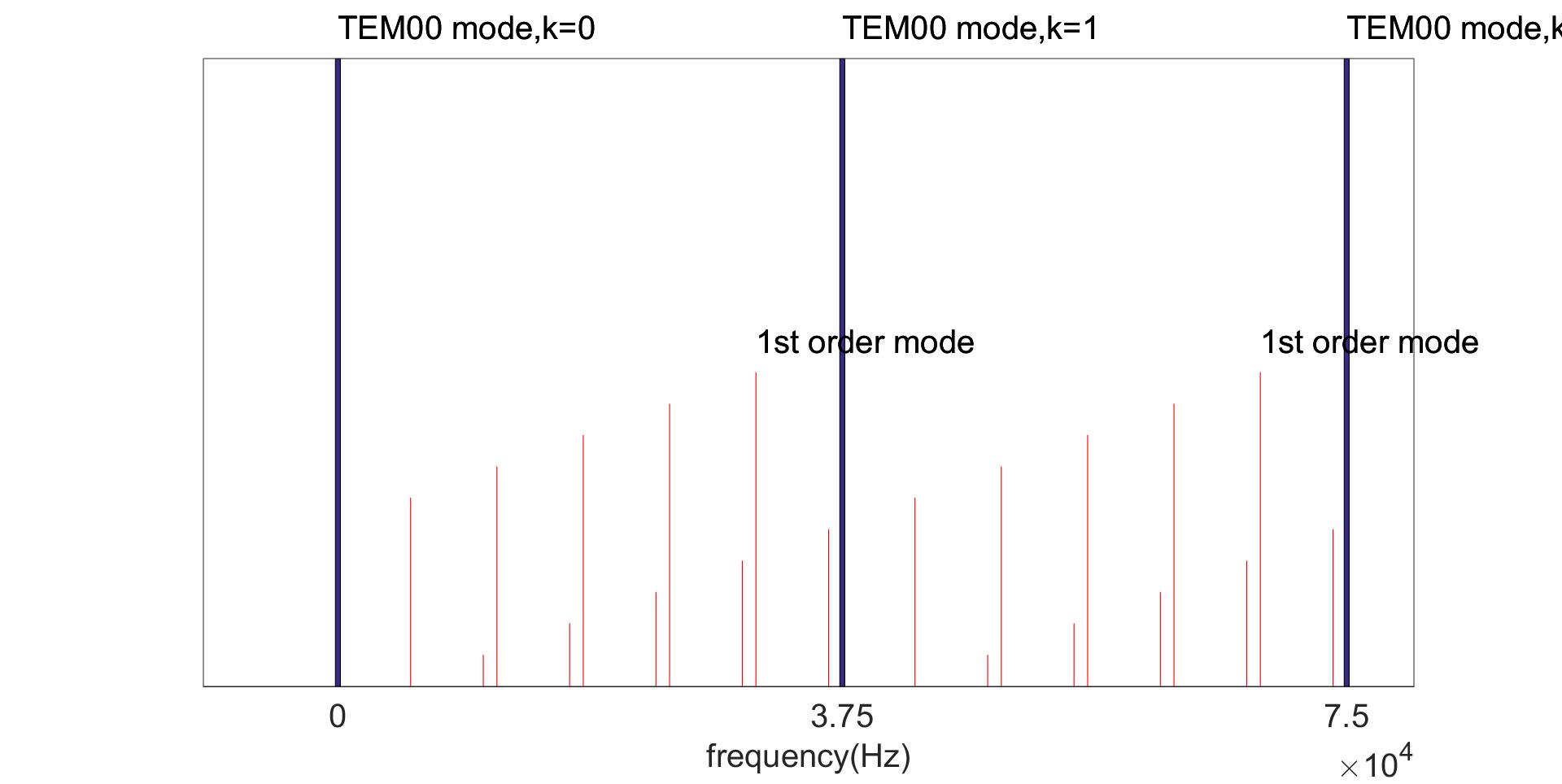}
				\caption{The cavity mode spectrum across 2 free spectral range.  The height of the lines represent different mode order.  The height decreases as the mode order increases.}
                \label{modespacing}
				\end{figure}

                In a Fabry-Perot cavity, the resonant frequency of higher-order modes are given as:
                \begin{equation}
                f_{nmp}=\frac{\pi k}{cL}+\frac{c}{L}(n+m+1)\arccos\left(\pm\sqrt{(1-\frac{L}{R_{1}})(1-\frac{L}{R_{2}})}\right),.
                \end{equation}
                The frequency spacing between the (n,m) transverse mode and the 00 mode is
                \begin{equation}
                 \omega_{0}-\omega_{mn}=k\cdot\mathrm{FSR}+\frac{c}{L}(m+n)\arccos\left(\pm\sqrt{(1-\frac{L}{R_{1}})(1-\frac{L}{R_{2}})}\right).
                 \end{equation} 
                Figure \ref{modespacing} shows a schematic diagram of mode structure of a cavity, up to $10^{th}$ order optical mode. Table \ref{stokes and anti-stokes} lists the mode gaps (frequency difference) between carrier mode and the Stokes/Anti-Stokes modes considered in this simulation.
  				\begin{table}[H]
                \centering
  			    \begin{tabular}{|c|c|c|}
			    \hline  
			    
			    optical mode order&$\omega_0 -\omega_{mn}[kHz]$  &  $\omega_{mn} -\omega_0[kHz]$\\
			    \hline
			    \multicolumn{3}{|l|}{k=0}\\
                \hline
                6&1&73\\
                7&7&67\\
                8&13&61\\
                9&20&54\\
                10&26&48\\
                \hline
                 \multicolumn{3}{|l|}{k=1}\\
                 \hline
                1&6&68\\
                2&12&62\\
                3&19&55\\
                4&25&49\\
                5&32&42\\
                6&38&36\\
                7&44&30\\
                8&51&23\\
                9&57&17\\
                10&64&10\\
                \hline
                \multicolumn{3}{|l|}{k=2}\\
                \hline
                1&43&31\\
                2&50&24\\
                3&56&18\\
                4&63&11\\
                5&69&5\\
                \hline
                \end{tabular}
                \caption{Stokes and anti-Stokes mode spacing in the cavity, assuming the carrier frequency at k=1.}
                \label{stokes and anti-stokes}
                \end{table}
                
                The strength of the scattering interaction is governed by a three-mode detuning parameter $\triangle\omega$ (Eq.~\ref{eq df}) and the ratio of $\triangle\omega/\delta$.  The half line-width (also called the relaxation rate) of the transverse mode $\delta$ is defined as 
           
                \begin{equation}
                \delta=\frac{\mathrm{FSR}}{2\mathcal{F}},
                \end{equation}
               where $\mathcal{F}$ is the finesse of the cavity, which is the same as Advanced LIGO. 
                Narrower linewidths represent smaller cavity loss and  higher Q-factor. For laser freuency $\omega$, the quality factor is defined as:
                \begin{equation}
                Q=\frac{\omega}{2\delta}. 
                 \end{equation}
                 
    \subsection{Overlap}
    \paragraph{ }
                The overlapping factor is given as
                \begin{equation}
                \Lambda=\frac{V(\int E^{00}(\vec{r})E^{*hom}(\vec{r})\mu_{\bot}(\vec{r})d\vec{r_{\bot}})^{2}}{\int |\mu(\vec{r})|^{2}d\vec{r_{\bot}}\int |E^{00}(\vec{r})|^{2}d\vec{r_{\bot}}\int |E^{hom}(\vec{r})|^{2}d\vec{r_{\bot}}},
                \end{equation}
                where $V$ is the volume of test mass. Mathematically, the spatial match between the carrier mode $E^{00}$, transverse mode $E^{hom}$, and the acoustic mode $\mu$ is defined by the integral $\int E^{00}(\vec{r})E^{*hom}(\vec{r})\mu_{\bot}(\vec{r})d\vec{r_{\bot}}$, where $\mu_{\bot}$ is the displacement vector of the acoustic mode normal to the test mass surface. The larger value of the integral, the better overlap is between the three modes. The integral $\int |\mu(\vec{r})|^{2}d\vec{r_{\bot}}$ represents the effective volume of the acoustic mode. 
                          
\begin{figure}[H]
                \centering
                \subfigure[Acoustic mode]{\label{overlap_a}
                \includegraphics[width=2.3in]{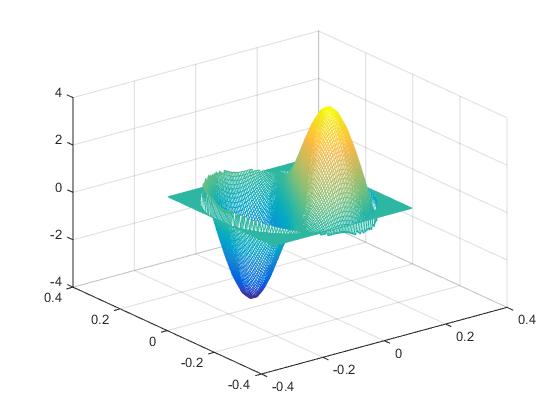}}
                \subfigure[Optical mode]{\label{overlap_b}
                 \includegraphics[width=2.3in]{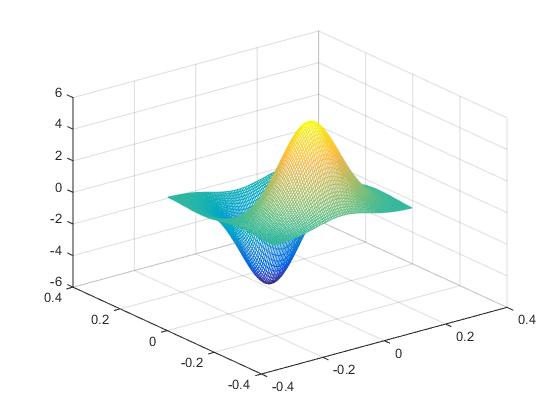}}
                 \caption{Mode shapes of one of the acoustic mode and optical 01 mode showing similarity in spatial distribution. }
                 \label{overlap}
                 \end{figure}

                Figure \ref{overlap} shows an example of  the shape of one of the acoustic mode and an optical 01 mode, which are very similar and for which the overlap will be high. .
             
                     In our analysis, we considered the worst case scenario by rotating the displacement of each acoustic mode every $\frac{\pi}{18}$ from 0 to $\pi$ to locate the best match orientation with the optical modes to obtain the maximal overlapping parameter $\Lambda$. The overlap of each acoustic mode to all the 21 optical modes up to $10^{th}$ order considered here are calculated.              
               
                
    \subsection{Parametric Gain}
                 
\begin{figure}[H]

\centering
                \includegraphics[width=4.7in]{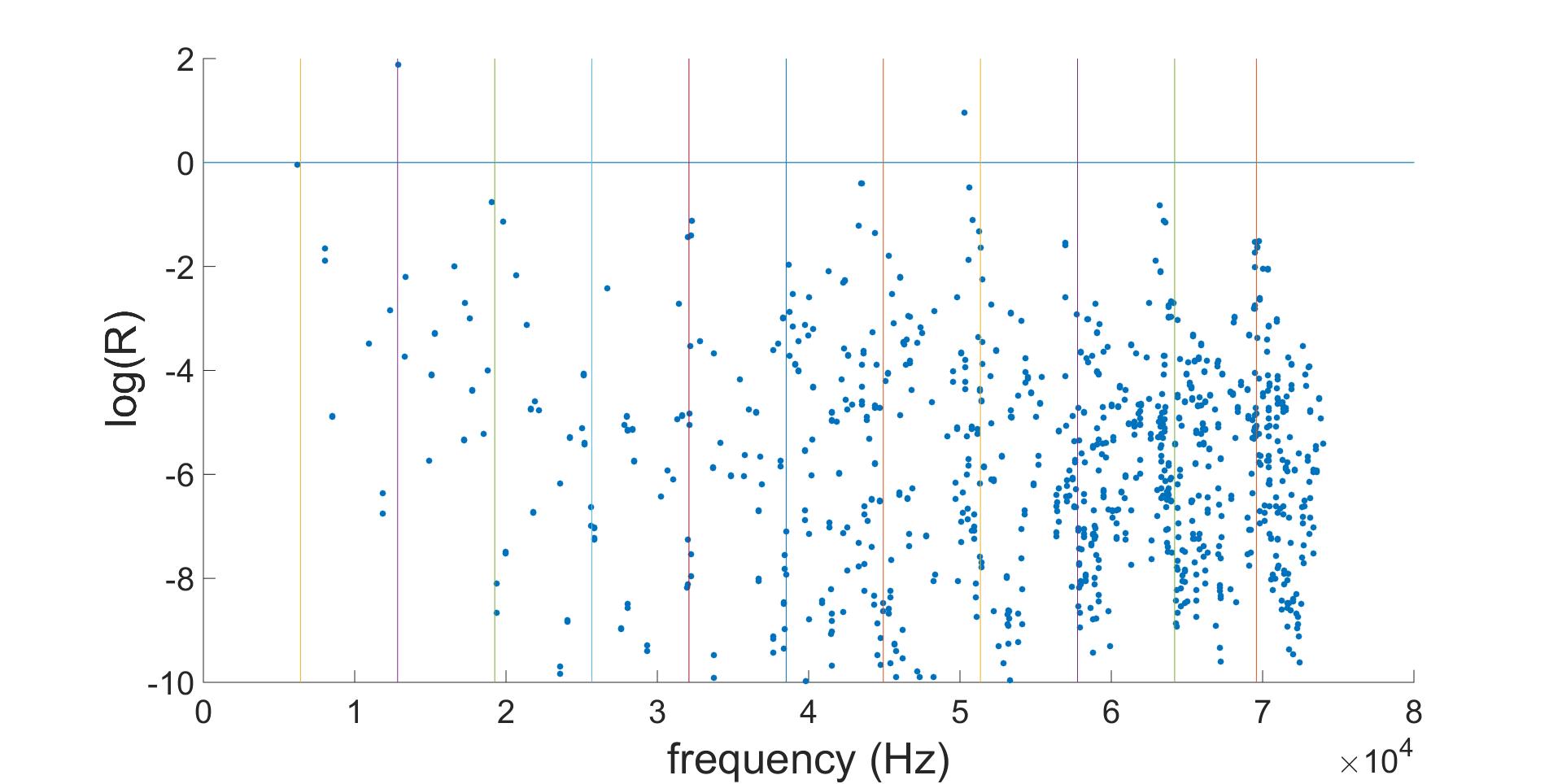}
                \caption{Parametric gain for different acoustic modes taking into account of one acoustic mode interacting with up to $10^{th}$ order optical modes, using parameters of Table \ref{properties}. The vertical lines represent the mode gap $\omega_{00}-\omega_{mn}$. }
                 \label{generalPI}
                \end{figure}

                Using the results of section 2.3, 2.4 and 2.5, we calculated the parametric gain for acoustic modes up to 74 kHz and up to $10^{th}$ order optical modes. Results is plot in Fig.~\ref{generalPI}. It can be seen that, at cavity power of 3MW, there are 1161 modes with $R>0$, of  which 2 modes have $R>1$, 4 modes with $R>0.1$, and 
               121 modes with $R>10^{-3}$. The maximum gain is 76. 

                 It can also be seen from Fig.~\ref{generalPI} that the frequency condition $(1+\triangle\omega^{2}/\delta^{2})^{-1}$ plays a big role in the parametric gain. 
                 Those acoustic modes having the frequencies near the mode gaps between the $TEM_{00}$ and $TEM_{mn}$ modes have higher parametric gain.

  			 
                 Thermal tuning of the RoC of the cavity (and thus changing the mode gaps of the optical cavity) was proposed  \cite{zhao2005parametric} to suppress parametric instability. However, due to the high acoustic mode density, changing the cavity mode gaps to suppress one acoustic mode instability may result in increasing other acoustic modes parametric gain.
                 
                  
				For cryogenic detectors, tuning of the RoC thermally may not be effective.  However, since there appears fewer unstable modes then in aLIGO, it would be relatively easy to use other methods such as electrostatic feedback or passive damper to suppress PI.  
                
\begin{figure}[H]
\centering
\includegraphics[width=4.7in]{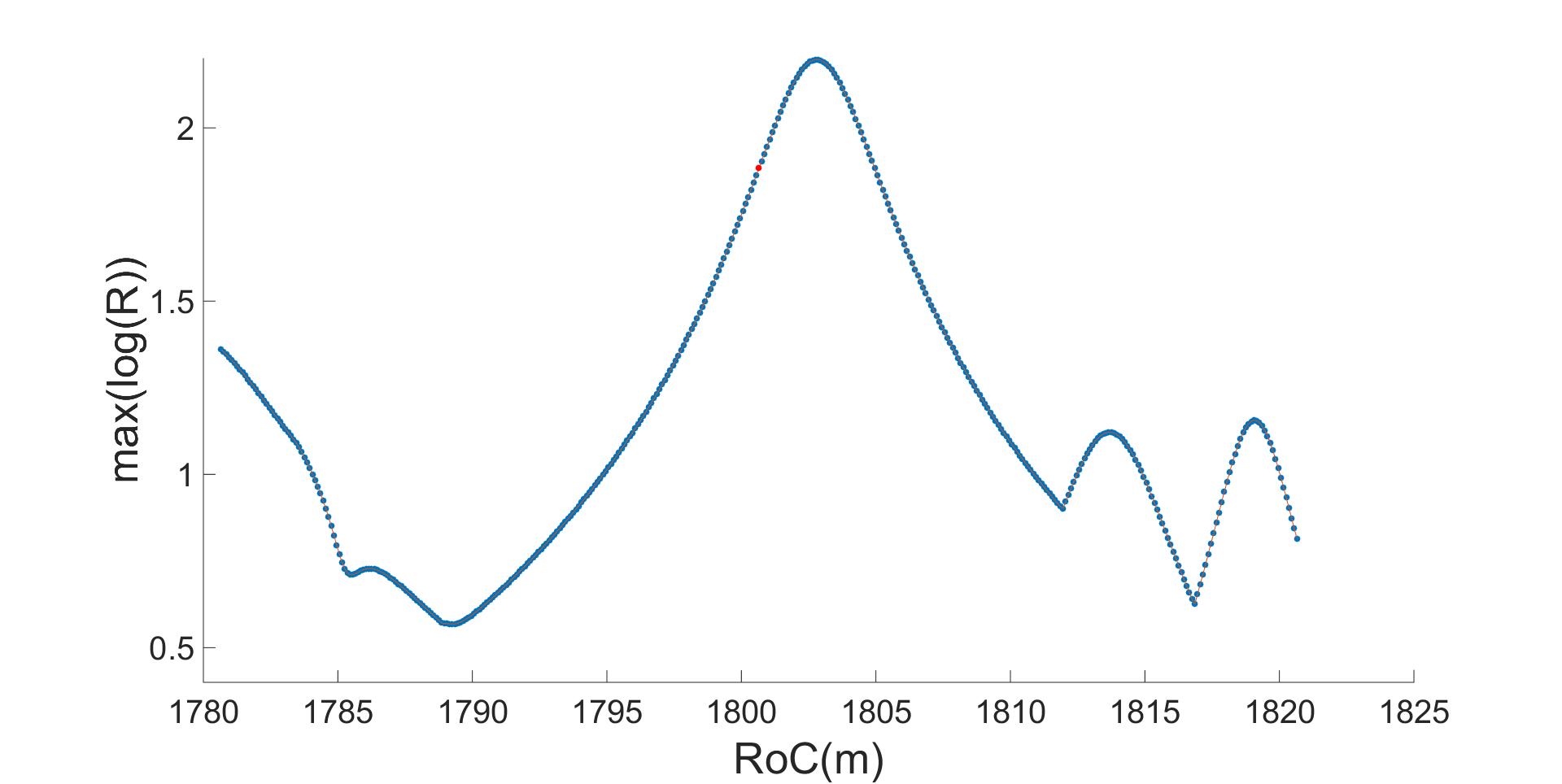}
\caption{Maximum parametric gain R as a function of the ITM RoC.}
\label{maxR}
\end{figure}

\begin{figure}[H]
\centering
\includegraphics[width=4.7in]{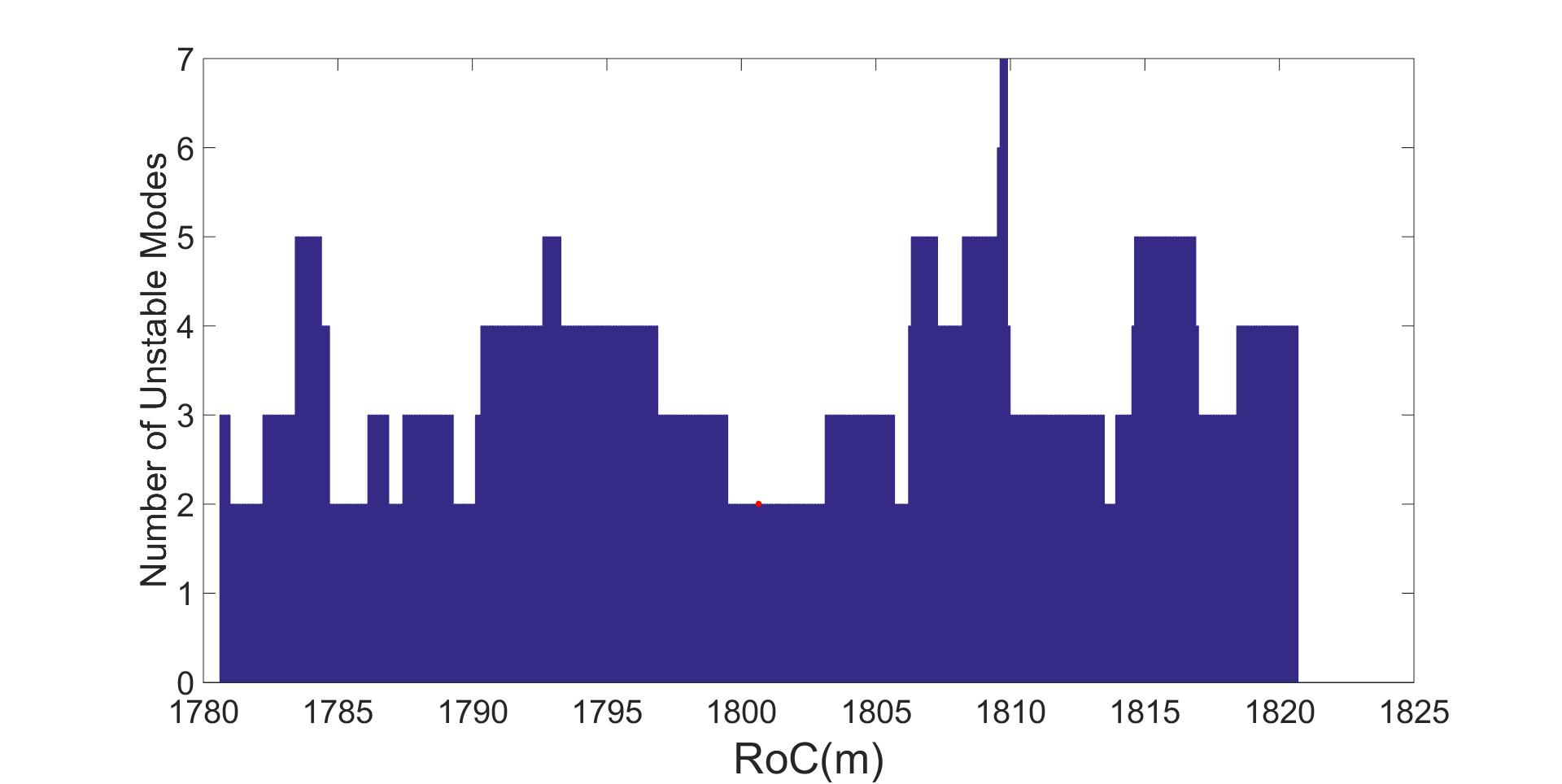}
\caption{Number of unstable modes as a function of the ITM RoC. }
\label{number of unstable modes}
\end{figure}

                It would be interesting to see if we could design the mirror RoC, so that there can be fewer acoustic modes with $R>1$.  Fig. \ref{maxR} and Fig. \ref{number of unstable modes} shows the maximum parametric gain and the number of unstable modes as a function of the RoC of the ITM. 

It can be seen that although there is nowhere that the parametric gain would be all below unity, there would be some "sweet regions" where both the number of the unstable modes and the maximum parametric gain are smaller. For example, at RoC $\approx$ 1790m, there would be less than 3 unstable modes with maximum R less than 4.  Even with the uncertainties in finite element modeling for acoustic mode frequencies (usually a few hundred to kHz different from measured values) \cite{strigin2008numerical},
Figs~\ref{maxR} and \ref{number of unstable modes}  provide useful information for choosing the favorable RoC of the test masses in regards to parametric instabilities. Besides, Fig \ref{maxR} shows that the maximum R comes from the same acoustic mode in the middle part ( RoC$\in [1790,1815]$ ) and then changes to another acoustic mode. The dominating acoustic mode is found out to be 12880 Hz interacting with second order optical mode. If we could damp this acoustic mode, the maximum parametric gain would be reduced dramatically. 

\begin{figure}[H]
\centering
\includegraphics[width=4.7in]{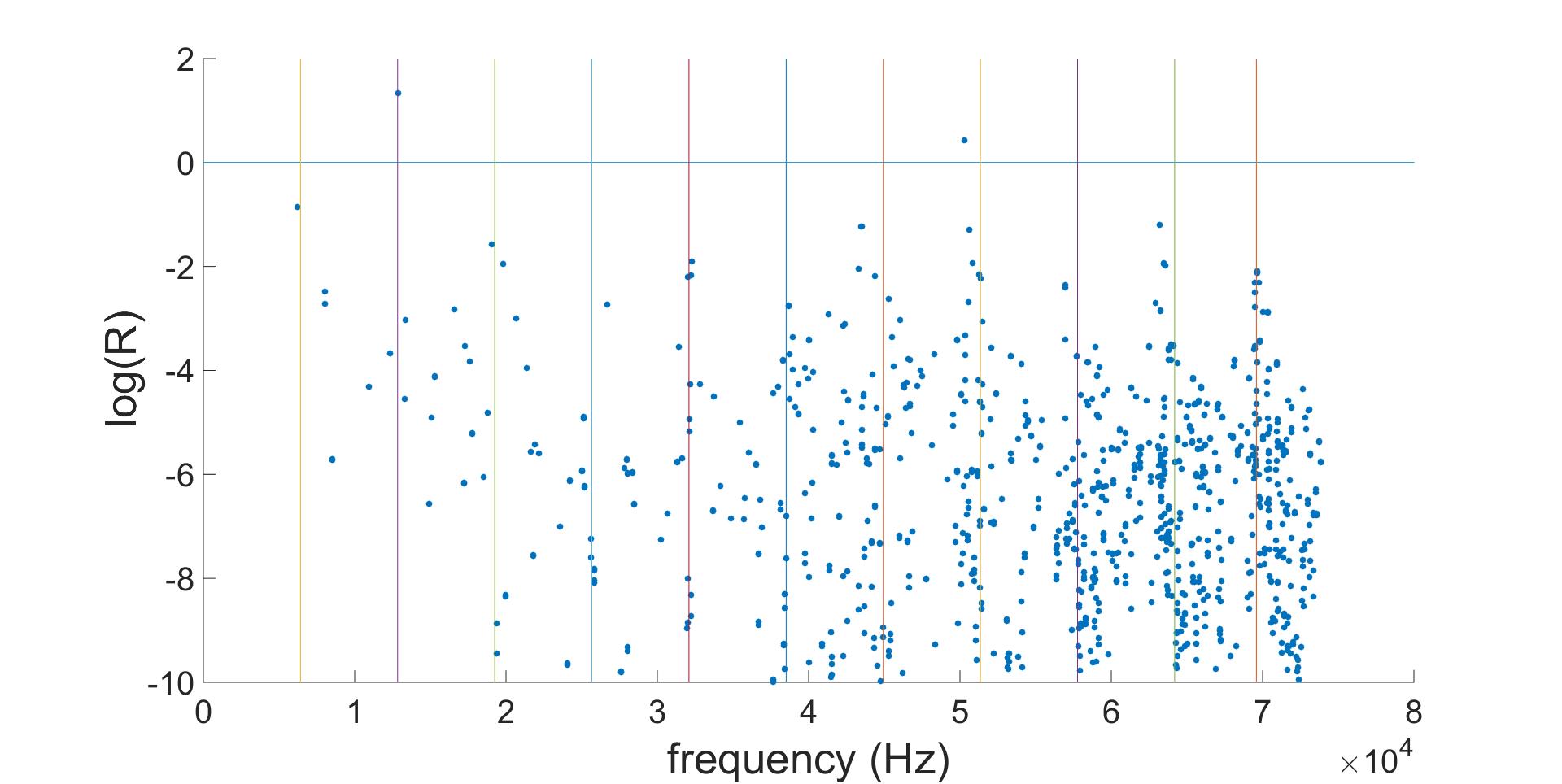}
\caption{Parametric gain for different acoustic modes with the same parameters as Fig. \ref{generalPI}, except cavity finesse $\mathcal{F}=3000$. }
\label{generalPIfinesse3000}
\end{figure}

\begin{figure}[H]
\centering
\includegraphics[width=4.7in]{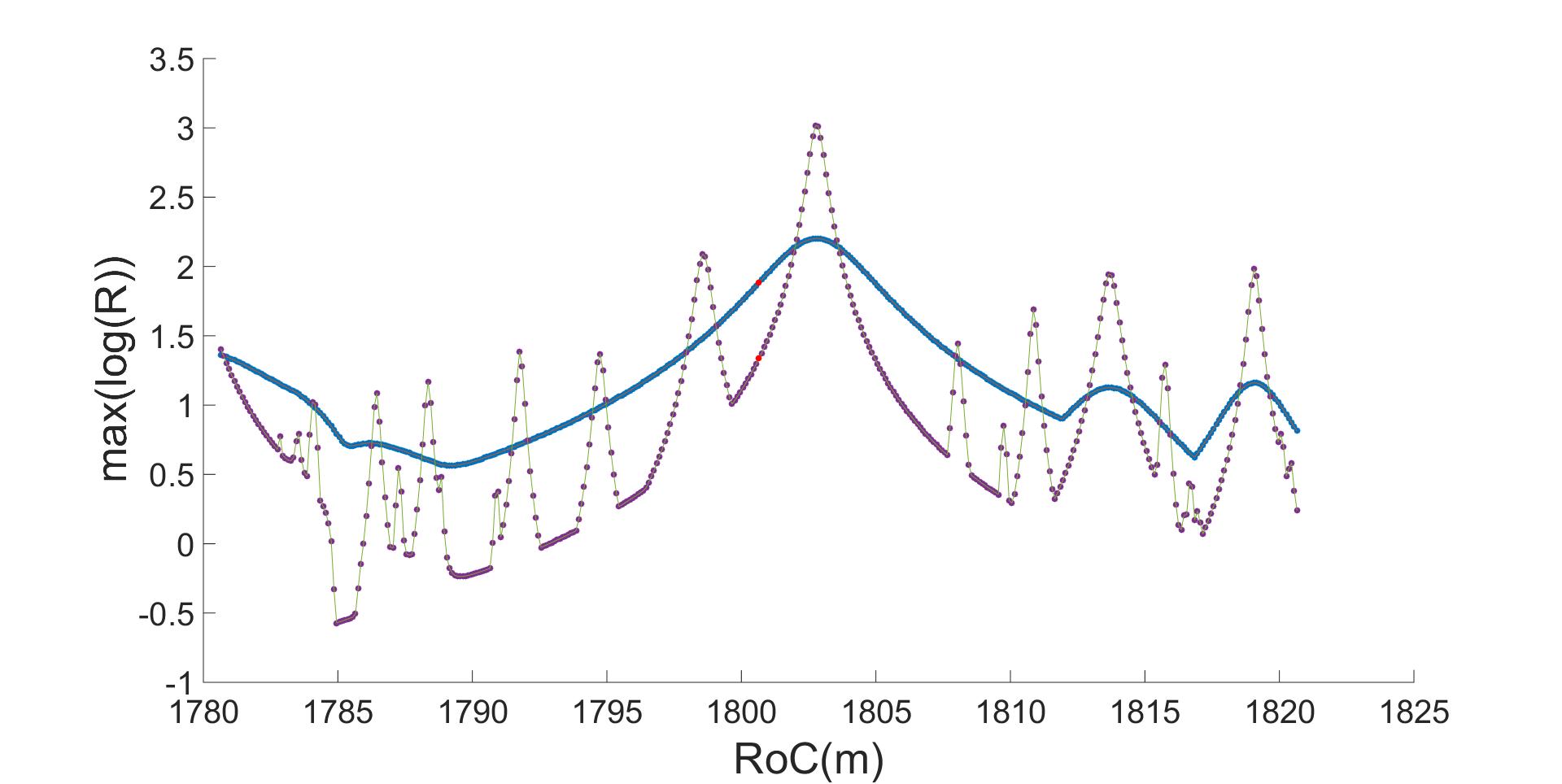}
\caption{Maximum parametric gain $R$ as a function of the ITM RoC for two arm cavity finesse, $\mathcal{F}=450$ (solid line) and $\mathcal{F}=3000$. }
\label{max R finesse 3000}
\end{figure}

\begin{figure}[H]
\centering
\includegraphics[width=4.7in]{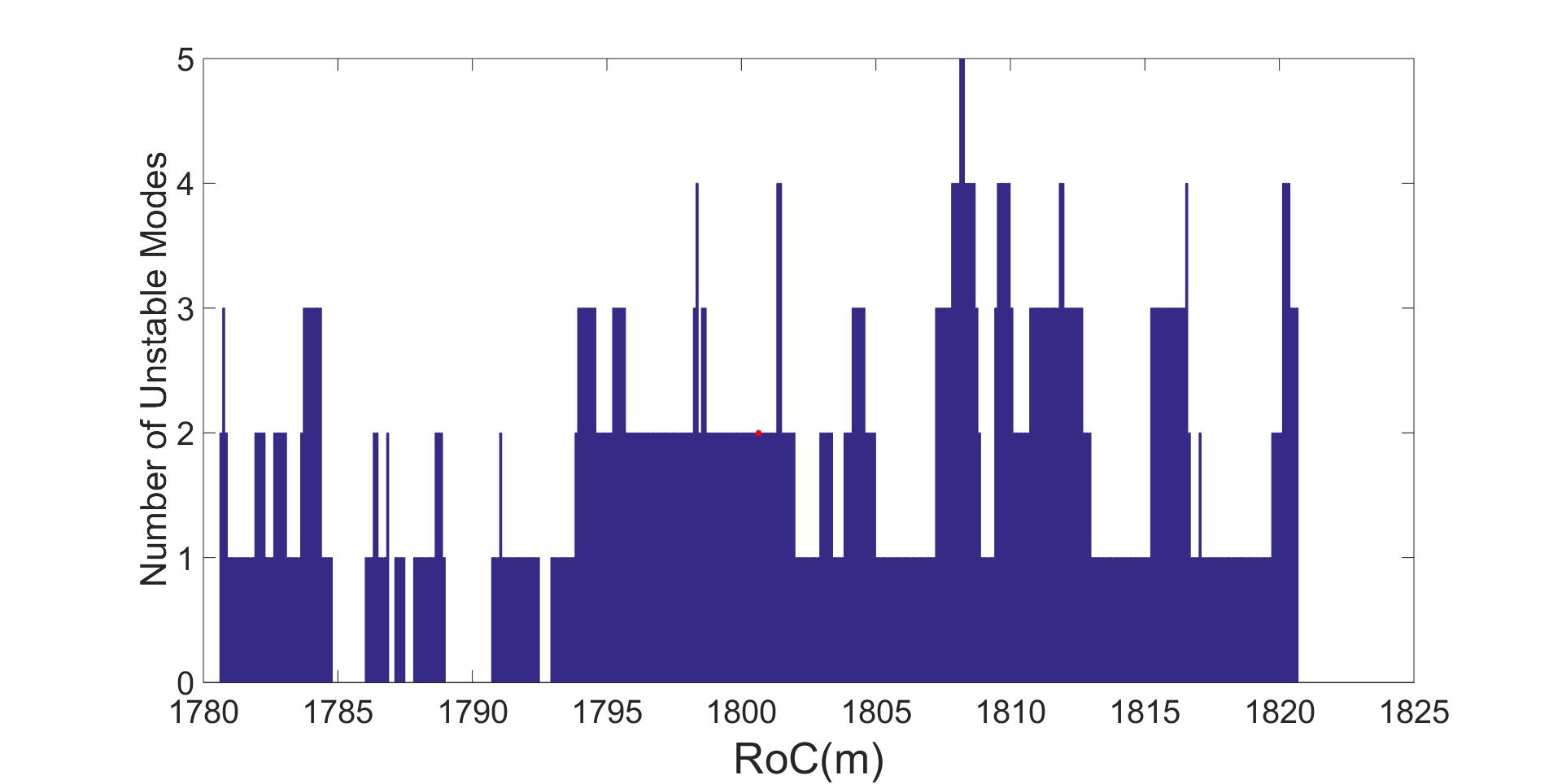}
\caption{Number of unstable modes as a function of the ITM RoC ($\mathcal{F}=3000$). }
\label{num of unstable modes finesse 3000}
\end{figure}

For silicon test masses, the optical absorption of large samples is still uncertain. It may be possible to avoid high absorbed power in the input test mass and beam splitter by using reduced circulating power in the power recycling cavity, compensated for by using  higher arm cavity finesse to maintain the same arm cavity circulating power (i.e. 3 MW). Here we have chosen a finesse of 3000 for the simulation to investigate any difference of parametric instability with this high cavity finesse. The results are shown in Figs. \ref{generalPIfinesse3000}, \ref{max R finesse 3000}
and \ref{num of unstable modes finesse 3000}.


It can be seen that for the particular RoCs in Table \ref{properties}, the unstable modes for the high finesse case is the same as the lower finesse case, with a smaller maximum gain $R=21$. From Figs.~\ref{max R finesse 3000} and \ref{num of unstable modes finesse 3000}, it is clear that due to the higher Q-factor and narrower linewidths of optical modes for high finesse cavity, the worst parametric gain is much higher than that of lower finesse, but with fewer unstable modes. It can also be seen that for the high finesse case, there would be several windows without instability, such as those near RoC $\sim 1785$m, RoC $\sim 1790$m. Moreover, there are more windows with only one unstable mode which will be easier to suppress.
\section{Conclusion} 

The result presented here is the first step in analyzing the parametric instability in the laser interferometer gravitational wave detector with silicon test masses. The single cavity model was used for simplicity.   In  practical configurations, the power and signal recycling cavities will change the parametric gain depending on the resonance condition of the high order mode involved inside the recycling cavities. As reference\cite{gras2010parametric} pointed out that any unstable modes present in the single cavity will be potential threats in a full interferometer. The first detailed analysis of parametric instability of Advanced LIGO\cite{zhao2005parametric} using a single cavity model predicted that there would be 7 unstable acoustic modes per each fused silica test mass, with parametric gain $R$ up to 7, at  circulating power of 800 kW. This means about a  total of 28 unstable modes in 4 test masses of the two cavity arms.  So far Advanced LIGO observed total $\sim$10 unstable modes, with the maximum parametric gain of the order of 10, operating  at $\sim 25\%$ of designed power.  This is consistent with the the single cavity simulation.  A more complicated simulation \cite{gras2010parametric} that took into consideration the recycling cavities and losses gave deeper insight of the parametric instability of the real system.  However, The simpler single cavity simulation is a very good indicator about the potential parametric instability problem of future design.


Assuming a Q-factor of $10^7$ for the acoustic modes of the test mass, we estimated that there would be $\sim 2$ unstable modes with maximum parametric gain of 76 for the Voyager blue design. This means that the potential parametric instability risk is roughly at the same level as the advanced LIGO detectors, at a much higher optical power level. The choice of a high sound velocity test mass material is advantageous in regards to parametric instabilities.

As designs are further developed, it will be necessary to consider all the effects mentioned above as well as  details of test mass structures such as flats on the barrel for suspensions, and the diffraction losses of high order modes.
\section*{Acknowledgment}

The authors would like to thank the support from LIGO scientific collaborations.  We also like to thank many useful discussion with Dr. Jiayi Qin, Dr. Xu Chen and Carl Blair.  This work is supported by Australian Research Council and is part of the Australian Consortium for Interferometric Gravitational wave Astronomy.

\section*{Reference}
\bibliography{BIB.bib}                 

\begin{thebibliography}{10}

\bibitem{abbott2016observation}
BP~Abbott, Richard Abbott, TD~Abbott, MR~Abernathy, Fausto Acernese, Kendall
  Ackley, Carl Adams, Thomas Adams, Paolo Addesso, RX~Adhikari, et~al.
\newblock Observation of gravitational waves from a binary black hole merger.
\newblock {\em Physical review letters}, 116(6):061102, 2016.

\bibitem{abbott2016gw151226}
BP~Abbott, R~Abbott, TD~Abbott, MR~Abernathy, F~Acernese, K~Ackley, C~Adams,
  T~Adams, P~Addesso, RX~Adhikari, et~al.
\newblock Gw151226: Observation of gravitational waves from a 22-solar-mass
  binary black hole coalescence.
\newblock {\em Physical Review Letters}, 116(24):241103, 2016.

\bibitem{aasi2015advanced}
Junaid Aasi, BP~Abbott, Richard Abbott, Thomas Abbott, MR~Abernathy, Kendall
  Ackley, Carl Adams, Thomas Adams, Paolo Addesso, RX~Adhikari, et~al.
\newblock Advanced ligo.
\newblock {\em Classical and Quantum Gravity}, 32(7):074001, 2015.

\bibitem{accadia2012virgo}
T~Accadia, F~Acernese, M~Alshourbagy, P~Amico, F~Antonucci, S~Aoudia, N~Arnaud,
  C~Arnault, KG~Arun, P~Astone, et~al.
\newblock Virgo: a laser interferometer to detect gravitational waves.
\newblock {\em Journal of Instrumentation}, 7(03):P03012, 2012.

\bibitem{belczynski2016first}
Krzysztof Belczynski, Daniel~E Holz, Tomasz Bulik, and Richard O’Shaughnessy.
\newblock The first gravitational-wave source from the isolated evolution of
  two stars in the 40--100 solar mass range.
\newblock {\em Nature}, 534(7608):512--515, 2016.

\bibitem{braginsky2001parametric}
VB~Braginsky, SE~Strigin, and S~Pr Vyatchanin.
\newblock Parametric oscillatory instability in fabry--perot interferometer.
\newblock {\em Physics Letters A}, 287(5):331--338, 2001.

\bibitem{zhao2005parametric}
C~Zhao, L~Ju, J~Degallaix, S~Gras, and DG~Blair.
\newblock Parametric instabilities and their control in advanced interferometer
  gravitational-wave detectors.
\newblock {\em Physical review letters}, 94(12):121102, 2005.

\bibitem{PhysRevLett.114.161102}
Matthew Evans, Slawek Gras, Peter Fritschel, John Miller, Lisa Barsotti, Denis
  Martynov, Aidan Brooks, Dennis Coyne, Rich Abbott, Rana~X. Adhikari, Koji
  Arai, Rolf Bork, Bill Kells, Jameson Rollins, Nicolas Smith-Lefebvre,
  Gabriele Vajente, Hiroaki Yamamoto, Carl Adams, Stuart Aston, Joseph
  Betzweiser, Valera Frolov, Adam Mullavey, Arnaud Pele, Janeen Romie, Michael
  Thomas, Keith Thorne, Sheila Dwyer, Kiwamu Izumi, Keita Kawabe, Daniel Sigg,
  Ryan Derosa, Anamaria Effler, Keiko Kokeyama, Stefan Ballmer, Thomas~J.
  Massinger, Alexa Staley, Matthew Heinze, Chris Mueller, Hartmut Grote, Robert
  Ward, Eleanor King, David Blair, Li~Ju, and Chunnong Zhao.
\newblock Observation of parametric instability in advanced ligo.
\newblock {\em Phys. Rev. Lett.}, 114:161102, Apr 2015.

\bibitem{gras2010parametric}
Slawomir Gras, Chunnong Zhao, DG~Blair, and Li~Ju.
\newblock Parametric instabilities in advanced gravitational wave detectors.
\newblock {\em Classical and Quantum Gravity}, 27(20):205019, 2010.

\bibitem{degallaix2007thermal}
J{\'e}r{\^o}me Degallaix, Chunnong Zhao, Li~Ju, and David Blair.
\newblock Thermal tuning of optical cavities for parametric instability
  control.
\newblock {\em JOSA B}, 24(6):1336--1343, 2007.

\bibitem{ju2008strategies}
L~Ju, DG~Blair, C~Zhao, S~Gras, Z~Zhang, P~Barriga, H~Miao, Y~Fan, and
  L~Merrill.
\newblock Strategies for the control of parametric instability in advanced
  gravitational wave detectors.
\newblock {\em Classical and Quantum Gravity}, 26(1):015002, 2008.

\bibitem{gras2008test}
S~Gras, DG~Blair, and L~Ju.
\newblock Test mass ring dampers with minimum thermal noise.
\newblock {\em Physics Letters A}, 372(9):1348--1356, 2008.

\bibitem{gras2009suppression}
S~Gras, DG~Blair, and C~Zhao.
\newblock Suppression of parametric instabilities in future gravitational wave
  detectors using damping rings.
\newblock {\em Classical and Quantum Gravity}, 26(13):135012, 2009.

\bibitem{fan2010testing}
YaoHui Fan, Lucienne Merrill, ChunNong Zhao, Li~Ju, David Blair, Bram
  Slagmolen, David Hosken, Aidan Brooks, Peter Veitch, and Jesper Munch.
\newblock Testing the suppression of opto-acoustic parametric interactions
  using optical feedback control.
\newblock {\em Classical and Quantum Gravity}, 27(8):084028, 2010.

\bibitem{miller2011damping}
John Miller, Matthew Evans, Lisa Barsotti, Peter Fritschel, Myron MacInnis,
  Richard Mittleman, Brett Shapiro, Jonathan Soto, and Calum Torrie.
\newblock Damping parametric instabilities in future gravitational wave
  detectors by means of electrostatic actuators.
\newblock {\em Physics Letters A}, 375(3):788--794, 2011.

\bibitem{zhao2015parametric}
Chunnong Zhao, Li~Ju, Qi~Fang, Carl Blair, Jiayi Qin, David Blair, Jerome
  Degallaix, and Hiroaki Yamamoto.
\newblock Parametric instability in long optical cavities and suppression by
  dynamic transverse mode frequency modulation.
\newblock {\em Physical Review D}, 91(9):092001, 2015.

\bibitem{Gras2015damper}
S.~Gras, P.~Fritschel, L.~Barsotti, and M.~Evans.
\newblock Resonant dampers for parametric instabilities in gravitational wave
  detectors.
\newblock {\em Phys. Rev. D}, 92:082001, Oct 2015.

\bibitem{punturo2010einstein}
M~Punturo, M~Abernathy, F~Acernese, B~Allen, Nils Andersson, K~Arun, F~Barone,
  B~Barr, M~Barsuglia, M~Beker, et~al.
\newblock The einstein telescope: a third-generation gravitational wave
  observatory.
\newblock {\em Classical and Quantum Gravity}, 27(19):194002, 2010.

\bibitem{instruments}
LIGO~Scientific Collaboration.
\newblock {\em Instrument Science White Paper}.
\newblock https://dcc.ligo.org/LIGO-T1400316/public, 2015.

\bibitem{dwyer2015gravitational}
Sheila Dwyer, Daniel Sigg, Stefan~W Ballmer, Lisa Barsotti, Nergis Mavalvala,
  and Matthew Evans.
\newblock Gravitational wave detector with cosmological reach.
\newblock {\em Physical Review D}, 91(8):082001, 2015.

\bibitem{blair2015next}
David Blair, Li~Ju, ChunNong Zhao, LinQing Wen, HaiXing Miao, RongGen Cai,
  JiangRui Gao, XueChun Lin, Dong Liu, Ling-An Wu, et~al.
\newblock The next detectors for gravitational wave astronomy.
\newblock {\em Science China Physics, Mechanics \& Astronomy}, 58(12):1--34,
  2015.

\bibitem{strigin2012effect}
SE~Strigin.
\newblock The effect of parametric oscillatory instability in a fabry-perot
  cavity of the einstein telescope.
\newblock {\em Optics and Spectroscopy}, 112(3):373--376, 2012.

\bibitem{cole2013tenfold}
Garrett~D Cole, Wei Zhang, Michael~J Martin, Jun Ye, and Markus Aspelmeyer.
\newblock Tenfold reduction of brownian noise in high-reflectivity optical
  coatings.
\newblock {\em Nature Photonics}, 7(8):644--650, 2013.

\bibitem{ju2006multiple}
L~Ju, S~Gras, C~Zhao, J~Degallaix, and DG~Blair.
\newblock Multiple modes contributions to parametric instabilities in advanced
  laser interferometer gravitational wave detectors.
\newblock {\em Physics Letters A}, 354(5):360--365, 2006.

\bibitem{VoyagerDesign}
Aidan Brooks Lisa Barsotti Brett Shapiro Brian Lantz David McClelland Eric
  Gustafson Valery Mitrofanov Dennis Coyne Koji Arai Norna Robertson Callum
  Torrie Christopher~Wipf Rana~Adhikari, Nicolas~Smith.
\newblock {\em LIGO Voyager Upgrade Conceptual Design}.
\newblock https://dcc.ligo.org/DocDB/0112/T1400226/007/VoyagerConcept-v7.pdf,
  2015.

\bibitem{strigin2008numerical}
SE~Strigin, DG~Blair, S~Gras, and SP~Vyatchanin.
\newblock Numerical calculations of elastic modes frequencies for parametric
  oscillatory instability in advanced ligo interferometer.
\newblock {\em Physics Letters A}, 372(35):5727--5731, 2008.

\end{thebibliography}
\end{document}